\documentclass[ioplppt,twocolumn,pra,superscriptaddress,amsmath,showpacs,tightenlines]{revtex4}

\usepackage{graphicx}
\usepackage{epsf}
\usepackage{amsmath}
\usepackage{amssymb}
\usepackage{array,multirow}
\usepackage{color}
\usepackage[cp1250]{inputenc}

\def\extra#1{{}}

\newcommand{\JOSAB}{{\em J. Opt. Soc. Am. B~}}
\newcommand{\Nature}{{\em Nature~}}
\newcommand{\NJP}{{\em New J. Phys.~}}
\newcommand{\PLA}{{\em Phys. Lett. A~}}
\newcommand{\PRA}{{\em Phys. Rev. A~}}

\newcommand{\PRL}{{\em Phys. Rev. Lett.~}}

\def \etal{{\em et al.}}

\newcommand{\tr}{\mathrm{tr}}

\begin{document}

\title{Quantum circuits for amplification of Kerr nonlinearity
via quadrature squeezing}

\author{Monika Bartkowiak}
\affiliation{Faculty of Physics, Adam
 Mickiewicz University, PL-61-614 Pozna\'n, Poland}
\affiliation{Department of Theoretical Physics and
 History of Science, The Basque Country University (EHU/UPV), 48080
 Bilbao, Spain}

 \author{Lian-Ao Wu}
  \affiliation{Department of Theoretical Physics and
 History of Science, The Basque Country University (EHU/UPV), 48080
 Bilbao, Spain}\affiliation{IKERBASQUE, Basque Foundation for Science, 48011 Bilbao, Spain}

  \author{Adam Miranowicz}
\affiliation{Faculty of Physics, Adam
 Mickiewicz University, PL-61-614 Pozna\'n, Poland}
 \date{\today}

\begin{abstract}
Phase shifts induced by the Kerr effect are usually very
small at the single-photon level. We propose two circuits for
enhancing the cross-Kerr phase shift by applying one- and
two-mode quadrature squeezing operators. Our results are
based on the vector coherent state theory and can be
implemented by physical operations satisfying the commutation
relations for generators of the generalized special unitary
group SU(1,1). While the proposed methods could be useful for
the realization of quantum optical entangling gates based on
Kerr nonlinear media at the single-photon level, they also
indicate a general alternative approach to enhance
higher-order nonlinearities by applying lower-order nonlinear
effects.
\end{abstract}


\maketitle


\section{Introduction}

The optical Kerr effect has been attracting considerable
interest in quantum state engineering (see, e.g.,
Refs.~\cite{HarocheBook,GerryBook}) for, e.g., self-focusing,
self-phase modulation, photon blockade (also referred to as
optical state truncation)~\cite{blockade1,blockade2} and
quantum nondemolition measurements~\cite{QND1,QND2} as
demonstrated in a number of experiments
(see~\cite{HarocheBook} for references). Moreover, the Kerr
effect can be used for a generation of nonclassical
light~\cite{Tanas03} including self-squeezed
light~\cite{KerrSqeezing} and macroscopic quantum
superpositions, i.e., the so-called Schr\"odinger
cat~\cite{Yurke86} and kitten~\cite{Miranowicz90} states.

The optical Kerr effect is also a potential resource for
performing deterministic photon interactions for quantum
information processing (see, e.g.,
Refs.~\cite{GerryBook,Milburn89}). Unfortunately, the Kerr
effect is usually very weak at the single-photon level.
Moreover, recent studies showed~\cite{Shapiro06,Gea10,Fan13}
that the phase noise in the cross-Kerr interaction of small
numbers of photons could be significant and, thus, could
preclude an effective implementation of entangling gates,
like the conditional phase (CPHASE) gate, at the
single-photon level. Nevertheless, recent experiments
demonstrate the possibility to effectively produce, control
and measure a non-zero conditional phase shift induced by a
cross-Kerr modulation for very weak light. For example,
Fushman~\etal~\cite{Fushman08} measured nonlinear Kerr-like
phase shifts of 0.05$\pi$  (9 degrees)  in a single quantum
dot coupled to a photonic crystal nanocavity at the
single-photon level. The maximum observed phase shift in this
report was equal to 0.16$\pi$ (28.8 degrees). The average
nonlinear cross–Kerr phase shifts of up to 20 degrees per
photon at the single-photon level was observed by Hoi
\etal~\cite{Hoi13} in their recent experiments with coherent
microwave radiation generated in superconducting circuits
based on Josephson junctions. By comparison, Matsuda
\etal~\cite{Matsuda09} measured the nonlinear Kerr phase
shifts of $\sim 10^{-7}$rad in optical fibres in single-shot
experiments at the single-photon level. The reported
nonlinear phase shift can be increased to $\sim 10^{-4}$ for
fibers of the same nonlinearity but with a reduced loss of 1
dB/km and flattened group-velocity
dispersion~\cite{Matsuda09}.

The effective Hamiltonian describing cross-Kerr interaction
between modes $a$ and $b$ can be given as~\cite{GerryBook}:
\begin{equation}
\hat{H}_{\rm Kerr}=\hbar\chi^{(3)}\hat n_{a}\hat n_{b},
\label{HamiltonianKerr}
\end{equation}
where $\chi^{(3)}$ is the (rescaled) third-order susceptibility of
the nonlinear medium, $\hat n_{a}=\hat a^{\dagger}\hat a$ and
$\hat n_{b}=\hat b^{\dagger} \hat b$ are the photon-number
operators given in terms of the annihilation ($\hat a$ and $\hat
b$) and creation ($\hat a^\dagger$ and $\hat b^\dagger$)
operators. We analyze photon-number qubits as superpositions of
vacuum and single-photon Fock states. Using an appropriate strong
cross-Kerr interaction, it is possible to perform the CPHASE gate
on two qubits in such a way that the states
$\vert00\rangle,\vert01\rangle,$ and $\vert10\rangle$ are
unchanged, but the two single-photon states gain some additional
phase $\delta$, i.e., $\vert11\rangle\rightarrow
e^{i\delta}\vert11\rangle.$ In particular, for $\delta=\pi$, the
CPHASE gate becomes the controlled-sign (CSIGN) gate, which is
equivalent (up to a unitary transformation) to the controlled-NOT
(CNOT) gate.

The main aim of our paper is to show how squeezing can be applied
to increase the cross-Kerr nonlinearity.

Squeezed light is a useful resource in high-precision metrology
and quantum information processing including quantum communication
(e.g., for quantum entanglement distribution) and quantum
cryptography (e.g., for secure quantum key
distribution)~\cite{BraunsteinBook}. The following values of
quadrature squeezing were experimentally observed in
continuous-wave optical fields: -9 dB~\cite{Takeno07}, -10 dB (-13
dB)~\cite{Vahlbruch07}, -11.5 dB~\cite{Mehmet10}, and -12.7
dB~\cite{Eberle10}. The value of -13 dB is the estimation of
squeezing achieved in the experiment~\cite{Vahlbruch07} after
correction for detector inefficiency, which results in a 5\%
improvement~\cite{Polzik08}. Recently, a few experiments with
superconducting circuits~\cite{Wilson11,Flurin12} have
demonstrated the possibility of obtaining much stronger squeezing
in microwave fields, even much exceeding -20 dB below the
shot-noise level~\cite{Delsing13}. Squeezing of light pulses,
which is more adequate for our circuits, is typically much weaker
than continuous-mode squeezing. Probably, the highest reported
experimental pulse-mode squeezing is only about -3 dB below the
shot-noise level: -3 dB~\cite{Wenger04}, -3.1
dB~\cite{Takahashi08}, and -3.2 dB~\cite{Eto07}.

Our amplification circuits are described in detail in the next
sections. We summarize our amplified Kerr shifts for the above
experimentally relevant squeezing values in table~I and
Conclusions.

\section{Circuit based on single-mode squeezing}

First we present a two-mode circuit for the amplification of
the phase shift induced by the nonlinear cross-Kerr effect
using one-mode squeezing operators. Our derivation is based
on the vector coherence theory (for a review,
see~\cite{Zhang90}).

Let us consider only a two-qubit subspace of the total
photon-number space and define qubit states with photon numbers
$0$ and $1$ as $\vert0\rangle$ and $\vert1\rangle=\hat
a^{\dagger}\vert0\rangle$, respectively. Therefore, in this
subspace used for quantum computation, we can introduce the
operator $\hat Z_{a}=2\hat n_{a}-1$, which has only two
eigenvalues equal to $1$ and $-1$, so that $\hat Z_{a}^{2}=1$.
This operator can be further used to construct one of the
generators of the SU(1,1) group,
\begin{equation}
\hat \Gamma _{3} = \frac {1}{2}(2\hat n_{b}+1)\hat Z_a= \frac
{1}{2}(4\underbrace{\hat n_{a}\hat n_{b}}_{\rm
Kerr\,effect}+2\hat n_{a}-2\hat n_{b}-1), \label{eq:gamma3}
\end{equation}
which is equivalent (up to some additional phase shift of both
qubits) to the Kerr effect described by
equation~(\ref{HamiltonianKerr}). In order to preserve the bosonic
commutation rules for the generators of SU(1,1):
\begin{eqnarray}
[\hat \Gamma_{1},\hat \Gamma_{2}] &=&  -i2\hat  \Gamma_{3},\nonumber \\
~[\hat \Gamma_{2},\hat \Gamma_{3}] &=&  i2\hat \Gamma_{1}, \nonumber \\
~[\hat \Gamma_{3},\hat \Gamma_{1}] &=&  i2\hat \Gamma_{2},
\label{eq:comm}
\end{eqnarray}
we construct the remaining generators as follows:
\begin{eqnarray}
\hat \Gamma_{1} & = & \frac {1}{2}(\hat b\hat b+\hat b^{\dagger}\hat b^{\dagger})Z_{a},\nonumber \\
\hat \Gamma_{2} & = & \frac {i}{2}(\hat b\hat b-\hat
b^{\dagger}\hat b^{\dagger}), \label{eq:gamma12}
\end{eqnarray}
where $\hat b$ and $\ \hat b^{\dagger}$ fulfill the standard
bosonic commutation relation. Using the vector coherent-state
theory, we find a configuration of operations, which need to
be performed on qubits to amplify the conditional phase shift
induced by the cross-Kerr effect. The vector coherent-state
theory is based on the fact that structural constants depend
only on the commutation relations of generators, but they are
independent of the dimensions of the representations of those
generators, for example of the group SU(1,1),
\begin{eqnarray}
\exp(i\alpha\hat \Gamma_{k})\exp(i\beta\hat \Gamma_{l})
=\exp(i\xi\hat \Gamma_{1}+i\theta\Gamma_{2}+i\zeta\hat
\Gamma_{3}),
\end{eqnarray}
where, for given $k,l$ and $\alpha,\beta$, the structural
constants $\theta$, $\xi$ and $\zeta$ are independent of the
dimension of $\hat\Gamma_{k}.$ The generators of the group
SU(1,1), which is noncompact and does not have any finite
unitary representation, can, however, be written in a simple
two-dimensional non-Hermitian representation as:
\begin{eqnarray}
\hat \Gamma_{1} & = & i\hat
\sigma_{2}=\left[\begin{array}{cc}
0 & 1\\
-1 & 0\end{array}\right],\nonumber \\
\hat \Gamma_{2} & = & -i\hat
\sigma_{1}=\left[\begin{array}{cc}
0 & -i\\
-i & 0\end{array}\right],\nonumber \\
\hat \Gamma_{3} & = & \hat \sigma_{3}=\left[\begin{array}{cc}
1 & 0\\
0 & -1\end{array}\right].\label{eq:generator}
\end{eqnarray}
According to equation~(\ref{eq:generator}), we can design the
following setup for enhancing Kerr nonlinearity:
\begin{eqnarray}
e^{i\theta_1\hat \Gamma_{2}}  e^{i\frac{\delta}{2}\hat
\Gamma_{3}} e^{i\theta_2\hat \Gamma_{2}}  e^{i\frac
{\delta}{2}\hat \Gamma_{3}} e^{i\theta_1\hat \Gamma_{2}}
=e^{i \gamma\hat \Gamma_{3}},\label{eq:su11}
\end{eqnarray}
where the coefficients $\theta_2$ and $\gamma$ are introduced
via the angles $\delta$ and $\theta_1$ as follows:
\begin{eqnarray}
\theta_2&=& {\rm arctanh}[-\cos\delta\tanh (2\theta_1)],\nonumber \\
\gamma&=&\arctan[\tan\delta\cosh( 2\theta_1)]. \label{params}
\end{eqnarray}
The above result can be obtained as follows:
\begin{eqnarray}
\hat V\left(\begin{array}{cc}
e^{\frac{i\delta}{2}} & 0\\
0 & e^{-\frac{i\delta}{2}}\end{array}\right) w
 \left(\begin{array}{cc}
1 &x\\
x & 1\end{array}\right)
\hspace{1cm} \nonumber\\
\times\left(\begin{array}{cc}
e^{\frac{i\delta}{2}} & 0\\
0 & e^{-\frac{i\delta}{2}}\end{array}\right) \hat V =
\left(\begin{array}{cc}
y & 0\\
0 & y^*
\end{array}\right), \label{N1}
\end{eqnarray}
where
\begin{eqnarray}
\hat V&=&\left(\begin{array}{cc}
\cosh\theta_1 & \sinh\theta_1\\
\sinh\theta_1 &
\cosh\theta_1\end{array}\right), \nonumber \\
y&=&\frac{w} {\cosh(2\theta_1)}[\cos\delta+i\cosh (2
\theta_1)\sin\delta],
\end{eqnarray}
$x=-\cos \delta \tanh (2\theta_1)$ and $w=1/\sqrt{1-x^{2}}$. In
equation~(\ref{N1}), the exponential functions of $\hat\Gamma_{k}$
are group elements, which can be given in a matrix representation.
Although there is no finite unitary representation of the group,
we have checked our result also numerically for spaces of
relatively large dimension (up to 100).
\begin{figure}
\includegraphics[width=.48\textwidth]{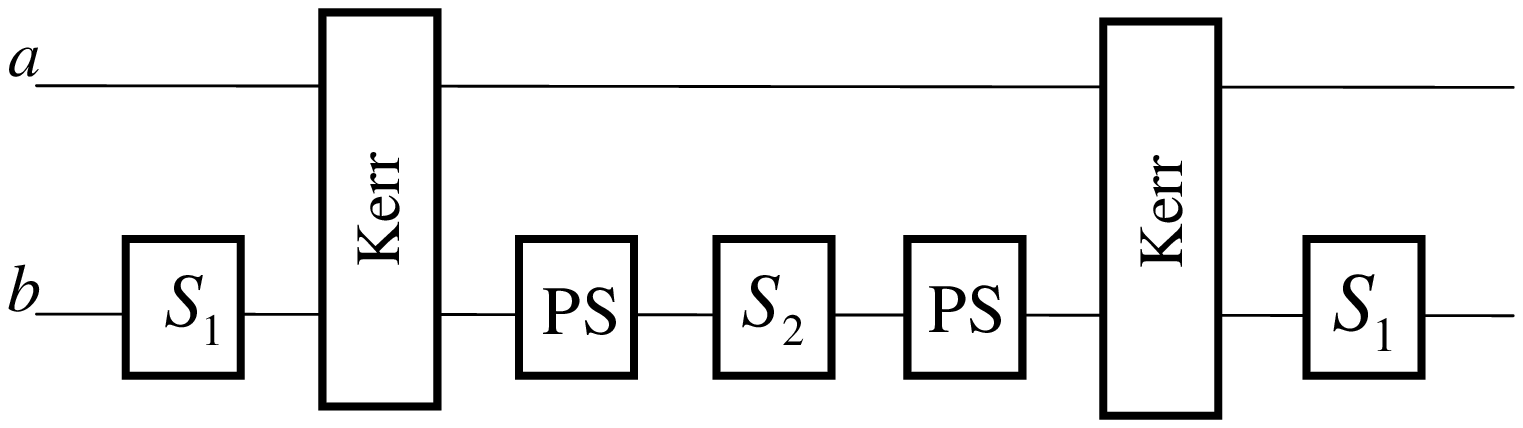}
\caption{A two-mode circuit for the amplification of the
cross-Kerr phase shift, where $\hat S_{1} = \hat
S_b(\theta_1)$ and $\hat S_{2}= \hat S_b(\theta_2)$ are the
single-mode squeezing operators for the parameters $\theta_1$
and $\theta_2$ given via equation~(\ref{params}), and PS are
the phase shifters.} \label{fig:circuit1}
\end{figure}
\begin{figure}
\includegraphics[width=.48\textwidth]{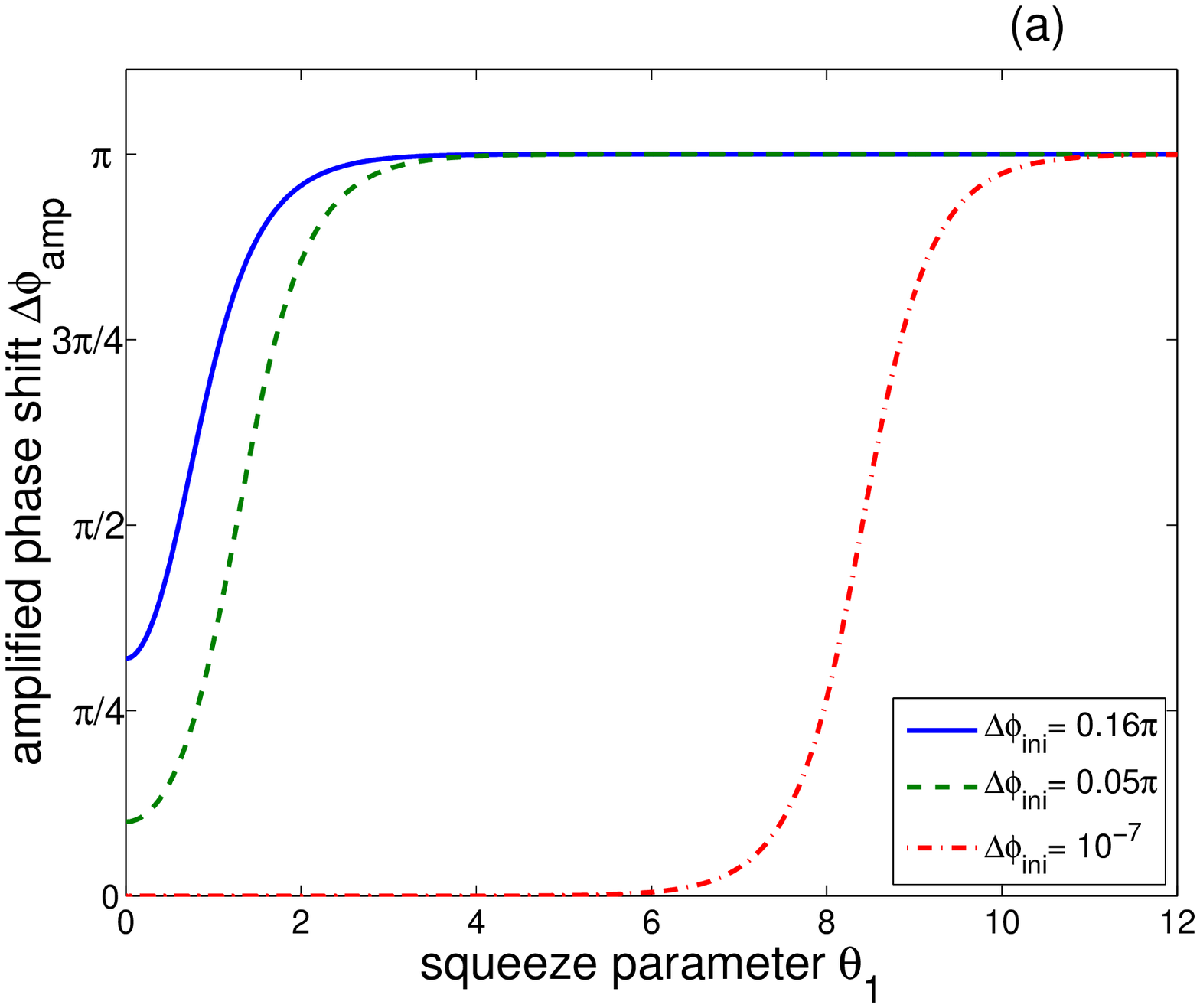}
\includegraphics[width=.48\textwidth]{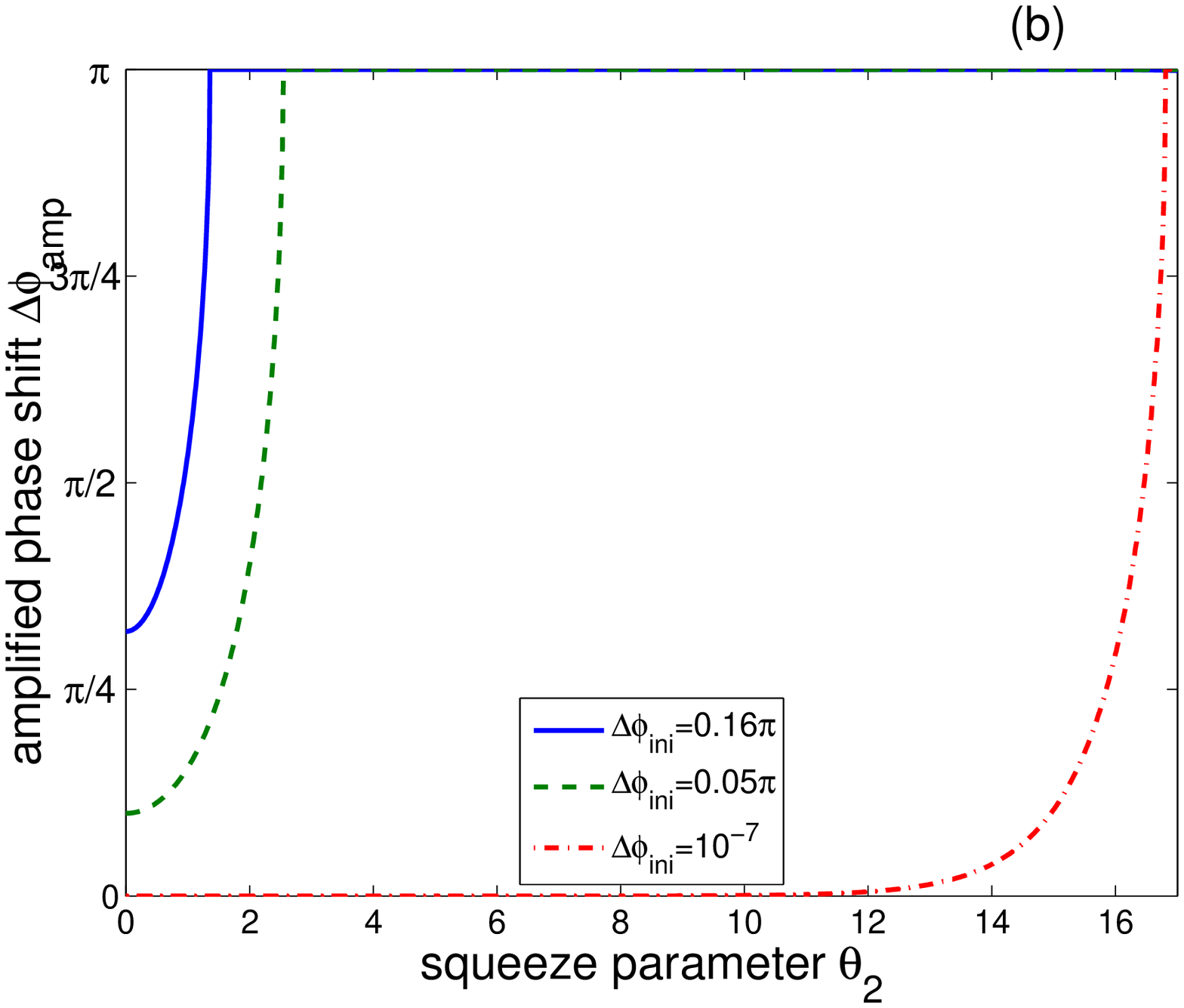}
\caption{(Color online) The amplified cross-Kerr phase shift
$\Delta\phi_{\rm amp}$ as a function of the squeezing
parameters (a) $\theta_1$ and (b) $\theta_2$ for various
values of the initial cross-Kerr phase shifts
$\Delta\phi_{\rm in}$.} \label{gamma}
\end{figure}

According to equation~(\ref{eq:su11}), it is possible to design
the following method for the Kerr-nonlinearity amplification:
\begin{eqnarray}
&\overbrace{\hat S_{b}(\theta_1)}^{\rm
squeezing}\overbrace{e^{i\frac{\delta}{2}(2\hat n_{a}\hat
n_{b}-\hat n_{b})}}^{\rm Kerr\&PS}\overbrace{\hat
S_{b}(\theta_2)}^{\rm squeezing}
\overbrace{e^{i\frac{\delta}{2}(2\hat n_{a}\hat n_{b}-\hat
n_{b})}} ^{\rm Kerr\&PS}\overbrace{\hat S_{b}(\theta_1)}^{\rm
squeezing}
\nonumber \\
&=\underbrace{e^{\frac{i}{2}(\gamma-\delta)(2\hat n_{a}-1)}}_{\rm
PS} \underbrace{e^{i\gamma(2\hat n_{a} \hat n_{b}-\hat
n_{b})}}_{\rm amplified\, Kerr\&PS},
\label{eq:amf_k}
\end{eqnarray}
where PS denotes (linear) phase shift in mode $b$. Based on
the vector coherence theory, we conclude that these relations
are valid in the whole ladder of Fock states for the mode
$b$. For the mode $a$, we restrict ourselves within the
subspace of the vacuum and single-photon states. The unitary
operation $\hat S_{b}(\theta_{1,2})$ [connected with the
exponents of $\hat\Gamma_2$ given in
equation~(\ref{eq:gamma12})] corresponds to the standard
single-mode (quadrature) squeezing operator \cite{GerryBook}:
\begin{eqnarray}
\hat S_k\equiv \hat S_{b}(\theta_k) & = &
\exp\Big[-\frac{\theta_k}{2}(\hat b\hat b-\hat b^{\dagger}
\hat b^{\dagger})\Big], \label{eq:sq}
\end{eqnarray}
where the squeezing parameter $\theta_k$ (with $k$=1,2) is
assumed to be real and extra minus corresponds to the
squeezing angle equal to $\pi$. This squeezing operator can
be implemented by a \emph{degenerate} parametric
down-conversion described by the interaction Hamiltonian
\begin{equation}
\hat{H}_{\rm pdc}=\frac{i}{2}\hbar\chi^{(2)}(\hat a^{\dagger}
\hat b\hat b-\hat a\hat b^{\dagger}
\hat b^{\dagger})
\label{HamiltonianPDC}
\end{equation}
in the strong classical pump limit, where the  operator $\hat
a$ is well approximated by a complex number $\alpha $.  The
interaction strength is proportional to the second-order
susceptibility, $\chi^{(2)}$, of the nonlinear medium. Thus,
this is a lower-order nonlinear process in comparison to the
Kerr effect given by equation~(\ref{HamiltonianKerr}). For
completeness, we note that this squeezing operator can be
also realized by higher-order nonlinear processes, e.g.,
described by $\chi^{(3)}$. The circuit, shown in figure~1 and
given by equations~(\ref{eq:su11}) and (\ref{eq:amf_k}), can
be compactly rewritten as:
\begin{eqnarray}
    \hat K(\Delta\phi_{\rm amp}) &=&  \hat P'
    \hat S_{1}\hat K(\Delta\phi_{\rm in})
    \hat P \hat S_{2} \hat P \hat K(\Delta\phi_{\rm in}) \hat S_{1},
\label{circuit1}
\end{eqnarray}
where the operators $\hat K(\Delta\phi)=\exp[i\Delta\phi \hat
n_a \hat n_b]$ describe the initial and amplified Kerr
effects, corresponding to the interaction strengths
\begin{equation}
\Delta\phi_{\rm in}=\delta,\quad \Delta\phi_{\rm
amp}=2\gamma, \label{strengths}
\end{equation}
respectively. Moreover in equation~(\ref{circuit1}), $\hat P=
\exp(-i\beta)$ and  $\hat P'= \exp(-i\beta')$ are linear phase
shifts with $\beta=\delta \hat n_b/2$ and
$\beta'=(\gamma-\delta)(\hat n_a-1/2)-\gamma \hat n_b$. The
right-hand side of equation~(\ref{circuit1}) is shown in figure~1
where, for simplicity, the less important gate $\hat P'$ is
omitted.

One can define the cross-Kerr effect amplification factor as the
ratio of the amplified, $\Delta\phi_{\rm amp}$, and initial,
$\Delta\phi_{\rm in}$, cross-Kerr phase shifts:
\begin{equation}
\kappa_{\rm amp}=\frac{\Delta\phi_{\rm amp}}
{\Delta\phi_{\rm in}}= \frac{2\gamma}{\delta}. \label{r1}
\end{equation}
Alternatively, one could define $\kappa'_{\rm
amp}=\kappa_{\rm amp}/2$, where factor 2 in the denominator
would count for two Kerr media used in this circuit (see
figure~1). In table~I, we calculated this amplification
factor for the best experimentally achieved values of the
squeezing parameters $\theta_2$ and the cross-Kerr phase
shifts $\Delta\phi_{\rm in}$.

\begin{table}
\caption{Examples of the amplified cross-Kerr phase shifts
$\Delta\phi^{(k)}_{\rm amp}$ and the amplification factors
$\kappa_{\rm amp}^{(k)}$ assuming experimental (see
references) and theoretical (those marked by [*]) values of
the squeezing parameter $\theta_2$ for various experimental
values of the initial nonlinear Kerr phase shifts: (1)
$\Delta\phi_{\rm in}^{(1)}\ll 1$ (e.g.,
$\Delta\phi^{(1)}_{\rm in}=10^{-7}$rad as measured in
reference~\cite{Matsuda09}), and (2) $\Delta\phi^{(2)}_{\rm
in}=0.05\pi=9^0$  and (3) $\Delta\phi^{(3)}_{\rm
in}=0.16\pi=28.8^0$  measured in reference~\cite{Fushman08}.
The squeezing parameter $\theta_1$ and $\Delta\phi^{(k)}_{\rm
amp}$ are calculated from equation~(\ref{params}).
Superscripts $p$ and $c$ refer to experiments with pulsed and
continuous-wave light, respectively.}
\begin{tabular}{l l r r r r r r}
\hline\hline
    $|\theta_2|$& reference                           & $|\theta_2|$  & $\kappa_{\rm amp}^{(1)}$& $\kappa_{\rm amp}^{(2)}$  & $\Delta\phi^{(2)}_{\rm amp}$& $\kappa_{\rm amp}^{(3)}$& $\Delta\phi^{(3)}_{\rm amp}$\\
    {[dB]}      &                                     & [rad]         &                    &                      & [deg]                       &                    & [deg] \\
\hline
    -3          &\cite{Wenger04,Takahashi08,Eto07}$^p$& 0.35          & 2.12               & 2.12                 & 19.1$^0$                    & 2.13               & 61.4$^0$\\
    -9          & \cite{Takeno07}$^c$                 & 1.04          & 3.17               & 3.19                 & 28.7$^0$                    & 3.46               &99.70$^0$\\
    -10         & \cite{Vahlbruch07}$^c$              & 1.15          & 3.48               & 3.51                 & 31.6$^0$                    & 3.95               &113.8$^0$\\
    -11.5       & \cite{Mehmet10}$^c$                 & 1.32          & 4.02               & 4.08                 & 36.7$^0$                    & 5.26               &151.6$^0$\\
    -13         & \cite{Eberle10,Vahlbruch07}$^{c}$   & 1.50          & 4.69               & 4.78                 & 43.0$^0$                    & ---                &180.0$^0$\\
    -20         & $[*]$                               & 2.30          & 10.10              &11.60                 &104.4$^0$                    & ---                &180.0$^0$
\\ \hline \hline
\end{tabular}
\end{table}

As already emphasized, Kerr effect is usually very small at
the single-photon level, i.e., $\Delta\phi_{\rm in}\ll 1$.
Let us also assume that the squeezing parameter $\theta_1$ is
relatively small such that $\tan(2\Delta\phi_{\rm in})\cosh
(2\theta_1)\ll 1$. Then, by expanding
equations~(\ref{params}) in power series in
$\delta=2\Delta\phi_{\rm in}$ and keeping only the first
terms of these expansions, one finds that $\theta_2\approx
-2\theta_1$ and
\begin{equation}
\quad \Delta\phi_{\rm amp}\approx4\Delta\phi_{\rm
in}\cosh(2\theta_1),
\end{equation}
which results in the Kerr amplification factor
\begin{equation}
\kappa_{\rm amp}\approx 2\cosh(2\theta_1) \label{r2}
\end{equation}
being independent of $\Delta\phi_{\rm in}$. The enhancements
of the cross-Kerr phase shift vs. squeezing parameters
$\theta_1$ and $\theta_2$ are plotted in figure~2 for
experimental (initial) nonlinear phase shifts reported by
Matsuda \etal~\cite{Matsuda09} and Fushman
\etal~\cite{Fushman08}. As can be seen in figure~2 and
table~I, we obtain a significant enhancement of the Kerr
nonlinearity. As it turns out, when an appropriate squeezed
light goes through two Kerr crystals and phase shifters, the
Kerr nonlinearity can be amplified to a $\pi$ shift. Thus,
the CPHASE gate can be, in principle, deterministically
implemented by Kerr nonlinearity via the cross-phase
modulation if appropriately strong squeezing of light is
available.

\section{Circuit based on two-mode squeezing}

Here we present a three-mode circuit for the amplification of Kerr
effect based on two-mode squeezing as an extension of the two-mode
circuit of the former section.

The two-mode squeezing operator acting on modes $b$ and $c$
can be defined as~\cite{GerryBook}:
\begin{eqnarray}
\hat S_{bc}(\theta_1) & = & \exp[-\theta_1(\hat b\hat c-\hat
b^{\dagger} \hat c^{\dagger})], \label{eq:sq2}\end{eqnarray}
with the real squeezing parameter $\theta_1$. This two-mode
squeezing operator can be implemented by a
\emph{nondegenerate} parametric down-conversion in a
$\chi^{(2)}$-nonlinear medium as described by the interaction
Hamiltonian
\begin{equation}
\hat{H}'_{\rm pdc}=i\hbar\chi^{(2)}(\hat a^{\dagger}
\hat b\hat c-\hat a\hat b^{\dagger}
\hat c^{\dagger})
\label{HamiltonianPDC2}
\end{equation}
assuming that the strong pump mode is treated classically, i.e.,
$\hat a\approx \alpha$, in analogy to the \emph{degenerate} case
described by equation~(\ref{HamiltonianPDC}). Also analogously to
equations~(\ref{eq:gamma3}) and~(\ref{eq:gamma12}), the generators
of the group SU(1,1) can be written as:
\begin{eqnarray}
\hat\Gamma'_{1} & = & (\hat b\hat c+\hat b^{\dagger}
 \hat c^{\dagger})\hat Z_{a},\nonumber \\
\hat\Gamma'_{2} & = & i(\hat b\hat c-\hat b^{\dagger}
 \hat c^{\dagger}), \nonumber \\
\hat\Gamma'_{3} & = & (\hat n_{b}+\hat n_{c}+1)\hat
Z_{a}\end{eqnarray} with the same commutation relations as those
given by equation~(\ref{eq:comm}) for $\hat \Gamma_{k}$. Based on
equation~(\ref{eq:su11}) we can derive the following relation:
\begin{eqnarray}
&\overbrace{\hat S_{bc}(\theta_1)}^{\rm squeezing}
\overbrace{e^{i\frac{\delta}{2}(2\hat n_{a}\hat n_{b}-\hat
n_{b})}} ^{\rm
Kerr_{ab}\&PS}\overbrace{e^{i\frac{\delta}{2}(2\hat n_{a}
\hat n_{c}-\hat n_{c})}}^{\rm Kerr_{ac}\&PS}
\overbrace{\hat{S}_{bc}(\theta_2)}^{\rm squeezing} \nonumber
\\ &\times\, \overbrace{e^{i\frac{\delta}{2}(2\hat n_{a}\hat
n_{b}-\hat n_{b})}} ^{\rm
Kerr_{ab}\&PS}\overbrace{e^{i\frac{\delta}{2}(2\hat n_{a}
\hat n_{c}-\hat n_{c})}}^{\rm Kerr_{ac}\&PS}
\overbrace{\hat{S}_{bc}(\theta_1)}^{\rm squeezing} \nonumber \\
&=\underbrace{e^{i(\gamma-\delta)(2\hat n_{a}-1)}}_{\rm PS}
\underbrace{e^{i\gamma(2\hat n_{a}\hat n_{b}-\hat n_{b})}
e^{i\gamma(2\hat n_{a}\hat n_{c}-\hat n_{c})}}_{\rm
amplified\ Kerr\&PS}.\label{eq:amf_k2}\end{eqnarray}

\begin{figure}
\includegraphics[width=8.5cm]{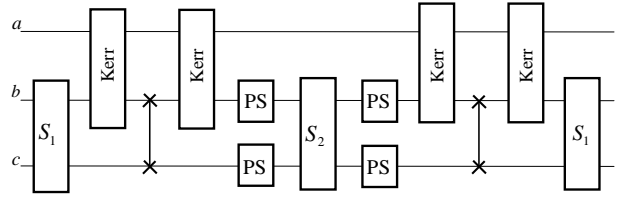}
\caption{A three-mode circuit for the amplification of the
cross-Kerr phase shift based on the two-mode squeezing
operators $\hat S_1\equiv \hat S_{bc}(\theta_1)$ and $\hat
S_2\equiv \hat S_{bc}(\theta_2)$. The SWAP gates are denoted
by lines connected with the symbols $\times$. The other
symbols are explained in figure~1.} \label{fig:circuit2}
\end{figure}
In analogy to the circuit of section II, this three-mode
circuit with the two-mode squeezing operators can be
compactly written in the form of equation~(\ref{circuit1}),
but for $\hat K=\hat K_{ab}\hat K_{ac}$ and $\hat P=\hat
P_b\hat P_c$, where $\hat K_{aj}=\exp(i \delta \hat n_a \hat
n_j)$ and $\hat P_{j}=\exp[-i (\delta/2) \hat n_j]$ for
$j=b,c,$ and $\hat{P}'= \exp(-i\beta')$ is the phase shift
with $\beta'=2(\gamma-\delta)(\hat n_a-1/2)-\gamma (\hat
n_b+\hat n_c)$. The final Kerr effect enhancement in this
circuit~is similar to the former circuit. Note that figure~3
shows a circuit composed of Kerr effect operators $\hat
K_{ab}$ applied in modes $a$ and $b$ solely. We have simply
applied the relation $\hat K=\hat K_{ab}\hat U^{bc}_{\rm
SWAP}\hat K_{ab} \hat U^{bc}_{\rm SWAP}$ in terms of the SWAP
operations $\hat U^{bc}_{\rm SWAP}$.

\section{Dissipation}

In order to include the effect of decoherence in our
circuits, we apply the beam-splitter model of losses, which
was developed in a general form by
Leonhardt~\cite{Leonhardt93} and d'Ariano~\cite{DAriano94}.

Realistic imperfect nonlinear elements (i.e., the squeezers
and Kerr elements) of the proposed circuits can be modelled
as the perfect ones followed by fictitious beam splitters
$\hat B_i$ (for $i=1,2,...$), as shown in figure~4. The
vacuum mode, assumed at one of the input ports of each beam
splitter, models the extra quantum noise caused by the
nonlinear effects.

Note that we ignore dissipation in phase shifters. In fact,
losses involved in linear optical elements (such as phase
shifters and beam splitters) are usually negligible in
comparison to those of realistic nonlinear optical elements.

As discussed in~\cite{Leonhardt93,Kiss95}, the model of
losses, based on a single-beam splitter, formally corresponds
to a dissipation described by the standard master equation
for a quiet reservoir (i.e., at zero temperature) as given by
\begin{eqnarray}
  \frac{d\hat\rho}{dt} &=& \tfrac12 (2 \hat b\hat \rho \hat b^\dagger
  -\hat b^\dagger\hat b\hat \rho-\hat \rho\hat b^\dagger\hat b).
\label{ME}
\end{eqnarray}
In this description, the formal time $t$ is simply related to
the reflectance as $R=1-\exp(-t)$ at the beam splitter.

The circuit shown in figure~4 (except the additional phase
shifter $\hat P'$) performs the following sequence of
operations
\begin{eqnarray}
    \hat K' &=&  \hat P' \hat B_5
    \hat S_{1}\hat B_4\hat B^{\prime}_4\hat K
    \hat P \hat B_3 \hat S_{2} \hat P \hat B_2\hat B^{\prime}_2
    \hat K \hat B_1 \hat S_{1},
\label{loss1}
\end{eqnarray}
where $\hat K=\hat K(\Delta\phi_{\rm in})$. The beam-splitter
transformations $\hat B'_i$ ($\hat B_i$) for the annihilation
operators $\hat a$ and $\hat b$ ($\hat v'_i$ and $\hat v_i$)
of the signal (vacuum) modes are given by
\begin{eqnarray}
  \hat B'_i &=& \exp[\theta_i (\hat a\hat v'_i{}^{\dagger}-\hat a^\dagger \hat v'_i)],
\nonumber \\
  \hat B_i &=& \exp[\theta_i (\hat b\hat v_i^\dagger-\hat b^\dagger \hat v_i)],
\label{BS}
\end{eqnarray}
respectively, where $\theta_i=\arccos(\sqrt{1-R_i})$ and
$R_i$ is the reflectance. Then, the output signal state $\hat
\rho_{\rm out}$ is given by
\begin{eqnarray}
  \hat \rho_{\rm out} &=& \tr_{e_{1},e_{2},e'_{2},e_{3},e_{4},e'_{4},e_{5}}
  [\hat K' (\hat \rho_{\rm in}\otimes
 \hat\rho_{\rm vac}) \hat K'^{\dagger}],
\label{loss2}
\end{eqnarray}
as obtained by tracing out the beam-splitter output modes $e_i$
and $e'_i$, which are lost to the environment. In this equation,
$\hat \rho_{\rm in}$ is the two-mode input signal state and
$\hat\rho_{\rm vac}=\hat\rho_{\rm vac}^{(e_1)}\otimes ... \otimes
\hat\rho_{\rm vac}^{(e_5)}$ with $\hat\rho_{\rm
vac}^{(e_i)}=(|0\rangle\langle 0|)_{e_i}$ are the vacuum modes. It
is rather inconvenient to directly apply equations~(\ref{loss1})
and~(\ref{loss2}) in numerical analysis. This would require
dealing simultaneously with nine-mode Hilbert spaces. Instead of
this, in our numerical simulation of losses, we have applied the
required operations sequentially as follows:
\begin{eqnarray}
 \hat \rho_1 &=& \tr_{e_{1}}[\hat B_1  (\hat S_1\hat \rho_{\rm in} \hat S_1^{\dagger}\otimes
 \hat\rho_{\rm vac}^{(e_1)})\hat B_1^{\dagger}],
 \nonumber \\
 \hat \rho_2 &=& \tr_{e_{2},e'_{2}}[\hat B_2 \hat B'_2( \hat K\hat \rho_1\hat K^{\dagger} \otimes
 \hat\rho_{\rm vac}^{(e_2)}\otimes
 \hat\rho_{\rm vac}^{(e'_2)}) \hat B_2'\!^{\dagger}\hat B_2^{\dagger}],
 \nonumber \\
 \hat \rho_3 &=&\hat P\, \tr_{e_{3}}[ \hat B_3  (\hat S_{2} \hat P\hat \rho_2\hat P^{\dagger} \hat S_{2}^{\dagger}\otimes
 \hat\rho_{\rm vac}^{(e_3)}) \hat B_3^{\dagger} ]\hat P^{\dagger},
  \nonumber \\
 \hat \rho_4 &=& \tr_{e_{4},e'_{4}}[\hat B_4 \hat B'_4 (\hat K\hat \rho_3 \hat K^{\dagger}\otimes
 \hat\rho_{\rm vac}^{(e_4)}\otimes
 \hat\rho_{\rm vac}^{(e'_4)}) \hat B_4'\!^{\dagger}\hat B_4^{\dagger}],
  \nonumber \\
 \hat \rho_{\rm out} &=& \hat P'\tr_{e_{5}}[\hat B_5 (\hat S_1 \hat \rho_4\hat S_1^{\dagger}\otimes
 \hat\rho_{\rm vac}^{(e_5)}) \hat B_5^{\dagger}]\hat P'\!^{\dagger}
\label{loss3}
\end{eqnarray}
according to the circuit shown in figure~4.

In order to compare the outputs of the perfect and lossy circuits,
we apply the Uhlmann-Jozsa fidelity defined
as~\cite{BengtssonBook}:
\begin{equation}
F(\hat \rho_{\rm ideal},\hat \rho_{\rm out})\equiv \Big[ {\rm
Tr}\Big(\sqrt{\sqrt{\hat \rho_{\rm ideal}}\hat \rho_{\rm
out}\sqrt{\hat \rho_{\rm ideal}}}\Big) \Big]^2.
\label{fidelity}
\end{equation}
In our case, $\hat \rho_{\rm out}$ is the output state of the
two-mode lossy circuit, given by equations~(\ref{loss2})
and~(\ref{loss3}), while $\hat \rho_{\rm ideal}$ is the ideal Kerr
state obtained by the application of the operators given by the
left- or right-hand side of equation~(\ref{circuit1}) to a given
initial state $\hat \rho_{\rm in}$. Note that the root fidelity
$\sqrt{F}$ is also sometimes referred to as fidelity (see,
e.g.,~\cite{NielsenBook}). Methods for measuring the fidelity and
its tight upper and lower bounds are described
in~\cite{Bartkiewicz13b}. If one of the states is pure, which can
be the ideal state $\hat \rho_{\rm ideal}=|\psi_{\rm
ideal}\rangle\langle\psi_{\rm ideal}|$, then the fidelity
simplifies to the straightforward expression $F=\langle \psi_{\rm
ideal}|\hat \rho_{\rm out} |\psi_{\rm ideal} \rangle$.

For simplicity, we present our numerical results only for the
cases when the losses in the signal modes in all the
squeezers (Kerr media), as shown in figure~4, are the same
and described by the reflectance $R_S\equiv R_1=R_3=R_5$
($R_K\equiv R_2=R'_2=R_4=R'_4$). If we set
$\theta_1=\Delta\phi_{\rm in}=1/2$, then the amplified
cross-Kerr phase shift is equal to $\Delta\phi_{\rm amp}=1.4$
for the ideal system.

Now we shortly discuss the results of our numerical simulations
for the two specific choices of the input state. Our first example
was calculated for an initial separable pure state, i.e.,
$|\psi_{\rm in}\rangle=|++\rangle,$ where $|+\rangle=
(|0\rangle+|1\rangle)/\sqrt{2}$. The perfect CPHASE gate should
transform this state into an entangled state $|\psi_{\rm
out}\rangle=(|00\rangle+ |01\rangle+
|10\rangle+e^{i\delta}|11\rangle)/2$. In contrast, the imperfect
amplifier generates a mixed state described by the fidelity
$F(R_K,R_S)\equiv F(\hat \rho_{\rm ideal},\hat \rho_{\rm out})$
depending on the chosen values of the losses (reflectances) $R_K$
and $R_S$. For example, we found $F(0.1,0.1)=0.74$,
$F(0.1,0)=0.82$, $F(0,0.1)=0.87$, and $F(0.2,0.2)=0.59$. We note
that $F(0,0)=1,$ as required for the ideal amplifier, and
$F(1,1)=|\langle 00| ++\rangle|^2=1/4$ for the amplifier absorbing
completely the input state.

Our second example is given for an initial entangled mixed
state, i.e., for the two-qubit Werner-like state defined by (see,
e.g.,~\cite{Miran04})
\begin{eqnarray}
  \hat \rho_{W} &=& p |\Phi^{+}\rangle \langle \Phi^{+}|
  +\frac{1-p}{4} \hat I,
\label{werner}
\end{eqnarray}
where $|\Phi^{+}\rangle=(|00\rangle+|11\rangle)/\sqrt{2}$ and
$\hat I$ is the identity operator. Moreover, we set $p=1/2$
(in general, $p\in [0,1]$). As in the former case we assume
$\theta_1=\Delta\phi_{\rm in}=1/2$, which  leads to
$\Delta\phi_{\rm amp}=1.4$ in the ideal system. The
calculated fidelities $F(R_K,R_S)$ as a function of the
losses (reflectances) $R_K$ and $R_S$ read as:
$F(0.1,0.1)=0.89$, $F(0.1,0)=0.929$, $F(0,0.1)=0.941$, and
$F(0.2,0.2)=0.81$. Note that, as for the former example,  it
holds $F(R_K,0)<F(0,R_S)$ for $R_S=R_K>0$. This can be
interpreted that our circuit is more sensitive to losses in
the Kerr media rather than those in the squeezers at least
for the analyzed values and states within our beam-splitter
model of losses. In addition, $F(0,0)=1,$ as expected for the
ideal case, and $F(1,1)=|\langle 00| \hat
\rho_W|00\rangle|^2=(1+p)/4=3/8$ for the amplifier absorbing
all the incident light.

Finally, we mention that a deeper analysis of decoherence in
our circuits should also include the effect of thermal
photons. This dissipation can be modelled by the full master
equation assuming that the thermal reservoir is at non-zero
temperature, which is a generalization of the quiet-reservoir
master equation, given by~(\ref{ME}). Such thermal effects
can be described (to some extent) by the beam-splitter model
of losses assuming that thermal photons, instead of the
vacuum mode, are at one of the input modes to fictitious beam
splitters. Preliminary studies show that our circuits are
strongly sensitive to thermal photons which is typical,
especially for nonlinear optical processes at the
single-photon level. We also note about the effects of mode
mismatch, which are important when dealing with
time-frequency overlaps of interfering light pulses. Such
mode-mismatch effects can affect efficiency of systems, which
can be revealed by applying a pulse-mode formalism, as
studied, e.g., in the related problem of a quantum scissors
system~\cite{Ozdemir02}. This infinite-mode formalism is
completely different from that applied here for a few modes
only. Our related theoretical~\cite{Ozdemir02} and
experimental studies~\cite{Lemr12} show that usually the
mode-mismatch problems can be effectively overcome in optical
experiments even at the single-photon level.

\begin{figure}
\includegraphics[width=.48\textwidth]{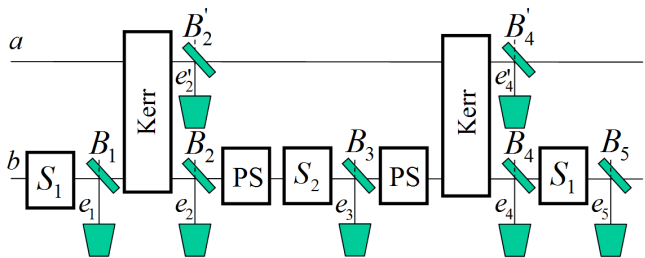}
\caption{(Color online) The application of the beam-splitter
model of losses in the circuit shown in figure~1. The output
signal modes are obtained by tracing out the modes $e_i$ and
$e'_i$, which are lost to the environment, as visualized by
green waste baskets. Broken lines at the second ports to the
beam splitters $\hat B_i$ ($i$=1,2...) denote the vacuum
modes, which model the extra quantum noise involved in the
nonlinear effects.} \label{fig4}
\end{figure}

\color{black}

\section{Conclusions}

We proposed two setups which can be used for enhancing the
phase shift in nonlinear cross-Kerr media, described by the
third-order nonlinear susceptibilities $\chi^{(3)}$,  by
applying a sequence of single-mode (or two-mode) squeezing
operators in media described by the second-order nonlinear
susceptibilities $\chi^{(2)}$. Our results are based on a
group-theoretical analysis. It is well known that entangling
gates, like controlled-sign (CSIGN) gate, cannot be
implemented deterministically using linear-optical elements
only (for a review see~\cite{Bartkowiak10}). Our approach
can, in principle, enhance the nonlinear phase shift to
180$^0$ at the single-photon level and thus enable a
deterministic implementation of the CSIGN gate if adequately
strong squeezed light source is available.

Our group-theoretical proposal can be implemented using various
systems exhibiting quadrature squeezing and cross-Kerr
nonlinearity. The predicted enhanced nonlinear phase shifts for
the experimentally observed initial nonlinear phase shifts and
generated squeezings are summarized in table~I.

We also studied dissipation in non-perfect circuits by
applying the beam-splitter model of losses. In particular, we
addressed the question how the Uhlmann-Jozsa fidelity,
between the outputs of the ideal and lossy systems, deviates
from 1 by the inclusion of losses.

Whilst we have proposed methods, which could be applied for an
implementation of quantum entangling gates using Kerr media at the
single-photon level, we have also shown an interesting general
idea to enhance higher-order nonlinear effects through other types
of lower-order nonlinear effects.

\noindent {\bf Acknowledgements.} The authors thank Nobuyuki
Matsuda and Patrick Leung for discussions. M.B. acknowledges
a scholarship from the Adam Mickiewicz University to stay at
the Basque Country University. This work was supported by the
Polish National Science Centre under grants
DEC-2011/03/B/ST2/01903 and DEC-2011/02/A/ST2/00305 the
Basque Government (Grant No. IT472-10) and the Spanish MICINN
(Project No. FIS2012-36673-C03-03).


\end{document}